\documentclass[twocolumn]{aastex631}

\usepackage{amsmath}
\usepackage{gensymb}

\shorttitle{Observations of \sn{} with \nustar}
\shortauthors{Andreoni et al.}
\graphicspath{{./}{figures/}}

\newcommand{\nustar}{\textit{NuSTAR}}
\newcommand{\sn}{SN 2018hti}
\def\mr{\mathrm}

\begin{document}

\title{Hard X--ray Observations of the Hydrogen-poor Superluminous Supernova \sn{} with \nustar}


\author[0000-0002-8977-1498]{Igor Andreoni}
\altaffiliation{Neil Gehrels Fellow}
\affil{Joint Space-Science Institute, University of Maryland, College Park, MD 20742, USA}
\affil{Department of Astronomy, University of Maryland, College Park, MD 20742, USA}
\affil{Astrophysics Science Division, NASA Goddard Space Flight Center, Mail Code 661, Greenbelt, MD 20771, USA}
\email{andreoni@umd.edu}

\author{Wenbin Lu}
\affil{Department of Astronomy and Theoretical Astrophysics Center, University of California, Berkeley, CA 94720-3411, USA}
\affil{Department of Astrophysical Sciences, Princeton University, Princeton, NJ 08544, USA}

\author{Brian Grefenstette}
\affil{Cahill Center for Astronomy and Astrophysics, California Institute of Technology, Pasadena, California, USA}

\author{Mansi Kasliwal}
\affil{Cahill Center for Astronomy and Astrophysics, California Institute of Technology, Pasadena, California, USA}

\author{Lin Yan}
\affil{Caltech Optical Observatories, California Institute of Technology, Pasadena, CA 91125, USA}

\author{Jeremy Hare}
\altaffiliation{NASA Postdoctoral Program Fellow}
\affil{Astrophysics Science Division, NASA Goddard Space Flight Center, Mail Code 661, Greenbelt, MD 20771, USA}

\begin{abstract}
Some Hydrogen-poor superluminous supernovae are likely powered by a magnetar central engine, making their luminosity larger than common supernovae. Although a significant amount of X--ray flux is expected from the spin down of the magnetar, direct observational evidence is still to be found, giving rise to the ``missing energy" problem. Here we present \nustar{} observations of nearby \object{SN 2018hti} 2.4\,y (rest frame) after its optical peak. We expect that, by this time, the ejecta have become optically thin for photons more energetic than $\sim 15$\,keV. No flux is detected at the position of the supernova down to $F_{\rm{10-30keV}} = 9.0\times 10^{-14}$\,erg\,cm$^{-2}$\,s$^{-1}$, 
or an upper limit
of $7.9 \times 10^{41}$\,erg\,s$^{-1}$ at a distance of 271\,Mpc.
This constrains the fraction of bolometric luminosity from the putative spinning down magnetar to be $f_{\rm X} \lesssim 36\%$ in the 10--30\,keV range in a conservative case, $f_{\rm X} \lesssim 11\%$ in an optimistic case.

\end{abstract}

\keywords{Transient sources (1851) --- X-ray transient sources (1852) --- Supernovae (1668)}


\section{Introduction} \label{sec:intro}

Explosions from the core collapse of massive stars generate supernovae (SNe) with a broad range of peak luminosities, but typically falling below $10^{43}\rm\, erg\,s^{-1}$. Superluminous supernovae (SLSNe) are a rare class of stellar explosions with peak luminosities $>7 \times 10^{43}$\,erg\,s$^{-1}$ \citep{gal-yam2019_SLSN_review}, which is $> 10 \times$ more luminous than typical core-collapse SNe. Optical surveys have shown that the volumetric rate of SLSNe is of the order of $1\%$ of the SN population \citep{Fremling2020}.

Interaction between the ejecta and circumstellar material is likely responsible for the large luminosity of Hydrogen-rich (Type II) SLSNe. However, hydrogen-poor SLSNe (Type I, or SLSN-I) likely involves a different mechanism, the leading candidate is a highly magnetized neutron star that continuously spins down and pumps energy into the ejecta \citep{kasen10_magnetar_model, woosley10_magnetar_model, metzger14_ionization_breakout}.

The magnetar model fits the SLSN-I light curves well up to $\sim100$ days from the explosion, when the spectral energy distribution (SED) peaks at optical/UV wavelengths. At later times, the spin-down luminosity of the magnetar exceeds the optical luminosity and the model needs to be modified to allow the majority of the spin-down energy to directly leak out of the ejecta. Some evidence of late-time excess $\sim 800$ days past explosion was found with deep optical observations for SN 2016inl \citep{Blanchard2021}. However, the anticipated large amount of leaked energy, about $10^{50}$ to $10^{51}\rm\, erg$, has so far gone undetected \citep{ Bhirombhakdi2018_SN2015bn_X, Margutti2018}.

Follow-up campaigns of nearby SLSNe \citep{ Bhirombhakdi2018_SN2015bn_X, Margutti2018} were conducted in the soft X-ray band with {\it XMM-Newton}, {\it Chandra}, and the {\it Neil Gehrels Swift Observatory}, placing deep constraints on the emission between 0.3\,keV and 10\,keV.
Only one faint counterpart was found for PTF12dam \citep{Margutti2018}, but the flux was consistent with the underlying star forming activity in the host galaxy and the X-rays may not be from the SLSN source. As we demonstrate in Equation \ref{eq:3}, the non-detection in soft X--rays may be explained by the large bound-free optical depth of the ejecta. In the high-energy $\gamma$-ray ($\gtrsim1\rm\, GeV$) band, \textit{Fermi} LAT observations placed constraints on the luminosity $L_{\rm GeV}\lesssim 10^{42}\rm\, erg/s$ on a timescale of a few years after the explosion \citep{renault-tinacci18_LAT_constraints}.
Thus, the question remains open: where is the missing energy? And what percentage of it is emitted in the X-ray band?

The Nuclear Spectroscopic Telescope Array \citep[\nustar;][]{Harrison2013} offers the possibility to address these outstanding questions by measuring the fraction of the missing energy emitted in the 3--79\,keV range a few years after the explosion when the ejecta have become optically thin in the hard X--ray band. 

Among the Hydrogen-poor SLSNe present in public catalogs (such as the Transient Name Server and the Open Supernova Catalog) in early 2021, we deemed \sn\ particularly promising to be detected with \nustar. \sn\ was discovered by the Asteroid Terrestrial-impact Last Alert System \citep[ATLAS;][]{ToDe2018} on 2018-11-02 at coordinates RA\,$=$\,03$^{\rm{h}}$40$^{\rm{m}}$53.760$^{\rm{s}}$; Dec\,$=$\,+11${\degree}$46$'$37.38$''$ (J2000). The transient was then classified as a SLSN-I \citep{2018TNSCR1719....1B} at redshift $z=0.0614$ \citep{Fiore2022}. Follow-up observations and modeling for \sn\ are presented in \cite{Lee2019, lin20_2018hti, Fiore2022}.

The paper is organized as follows. We present \nustar\ observations of \sn\ and data analysis method in \S\ref{sec:data analysis}, the analysis results in \S\ref{sec:results}, a discussion on the implications for the missing energy problem in \S\ref{sec:discussion}, and our conclusions in section \S\ref{sec:conclusion}. Times are UT throughout the manuscript.

\section{Observations and data analysis}
\label{sec:data analysis}

\sn{} was observed with \nustar{} in two epochs at the beginning of \nustar\ GO Cycle 7 (proposal 7264; P.I. Andreoni). The first epoch (ID 40701008001) started on 2021-07-01 16:34, with an exposure time of 101,690\,s. The second epoch (ID 40701008002) started on 2021-07-06 18:57, with an exposure time of 53,887\,s.

Data were reduced using HEASoft v.6.29 and the NuSTAR Data Analysis Software (NuSTARDAS) v.2.1.1, in particular the \texttt{nupipeline} and \texttt{nuproducts} routines. South Atlantic Anomaly (SAA) effects on the background were mitigated using the \texttt{nucalcsaa} routine.

The first epoch was significantly affected by solar activity outside the SAA. To mitigate the background increase, we extracted a light curve of an empty background region ($r=100''$), binned by 100\,s bins and we identified those time intervals in which the rate exceeded the median rate by $3\times$ the standard deviation of the rates.  We repeated this process twice and excluded the affected time frames from the ``good time interval" (GTI). After this operation, the effective exposure times became of 98.4\,ks (FPMA) and 98.8\,ks (FPMB). The total exposure time resulting from both epochs and both FPMA and FPMB instruments was $t_{\rm{exp}} = 302.4$\,ks. 

The upper limits on the flux presented below (\S\ref{sec:results}) were obtained using the X--ray spectral fitting package \texttt{XSpec} \citep{Arnaud1996xspec} of the HEAsoft software suite \citep{2014ascl.soft08004N}.

\section{Results}
\label{sec:results}

\begin{figure}
    \centering
    \includegraphics[width=0.49\columnwidth]{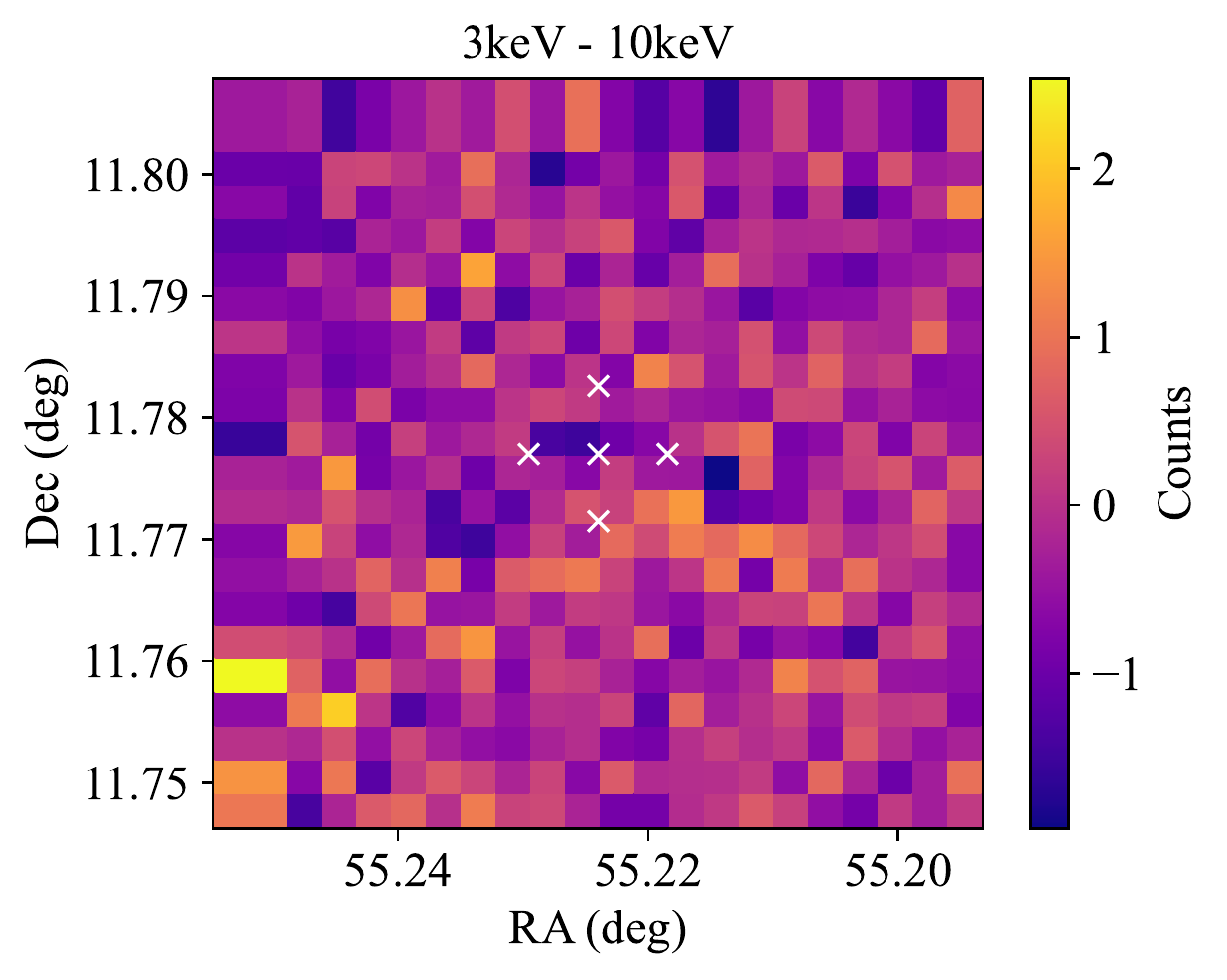}
    \includegraphics[width=0.49\columnwidth]{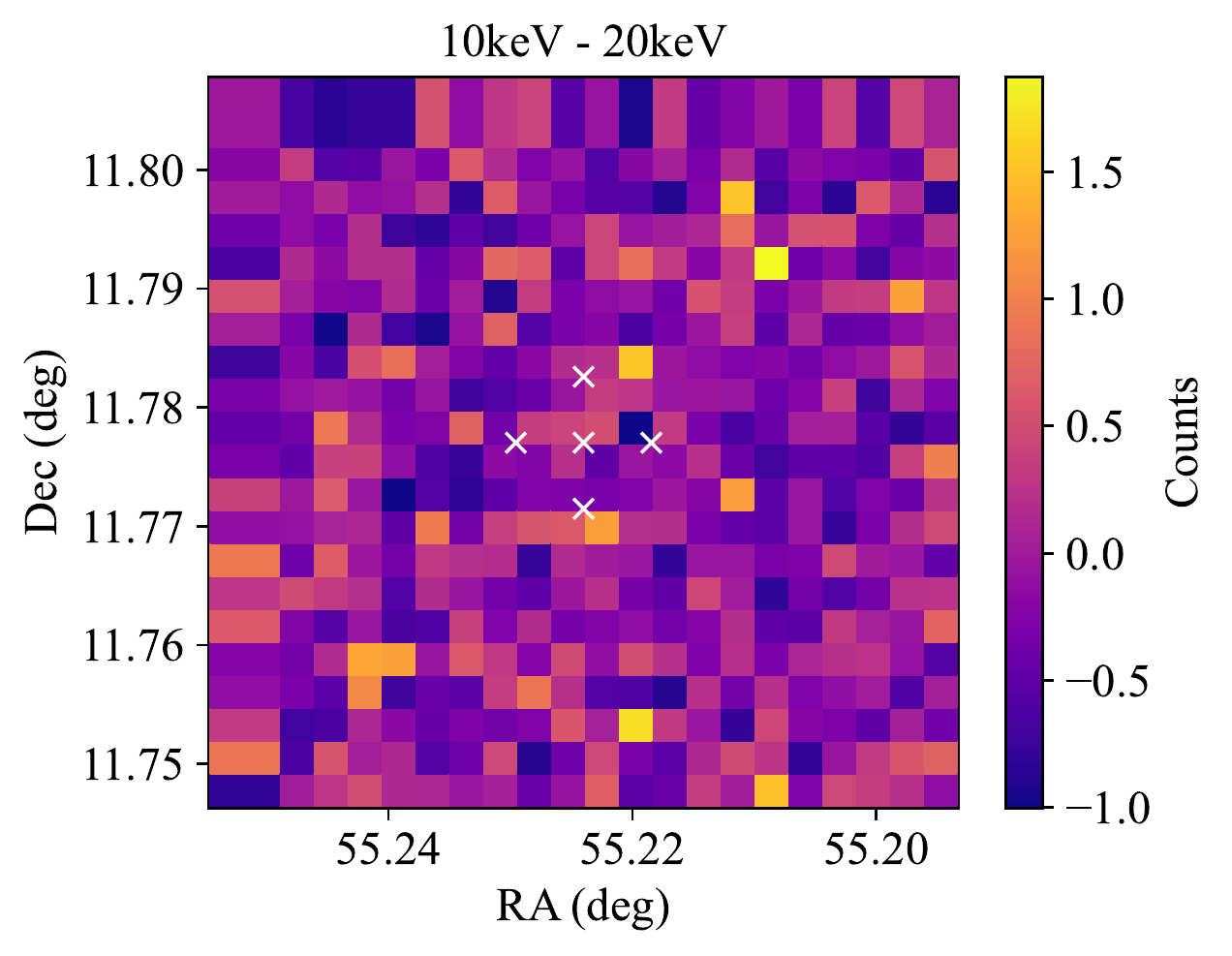}
    \includegraphics[width=0.49\columnwidth]{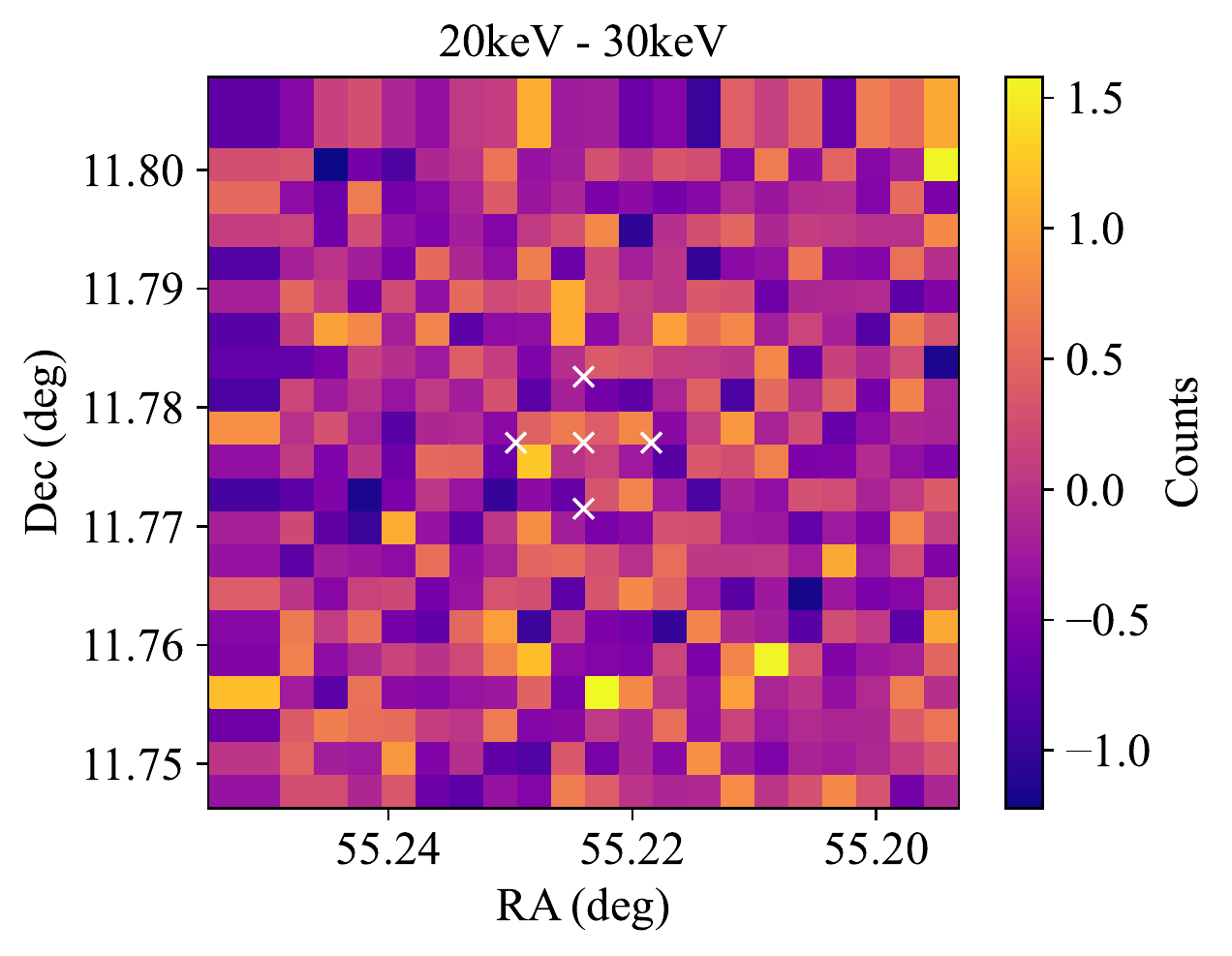}
    \includegraphics[width=0.49\columnwidth]{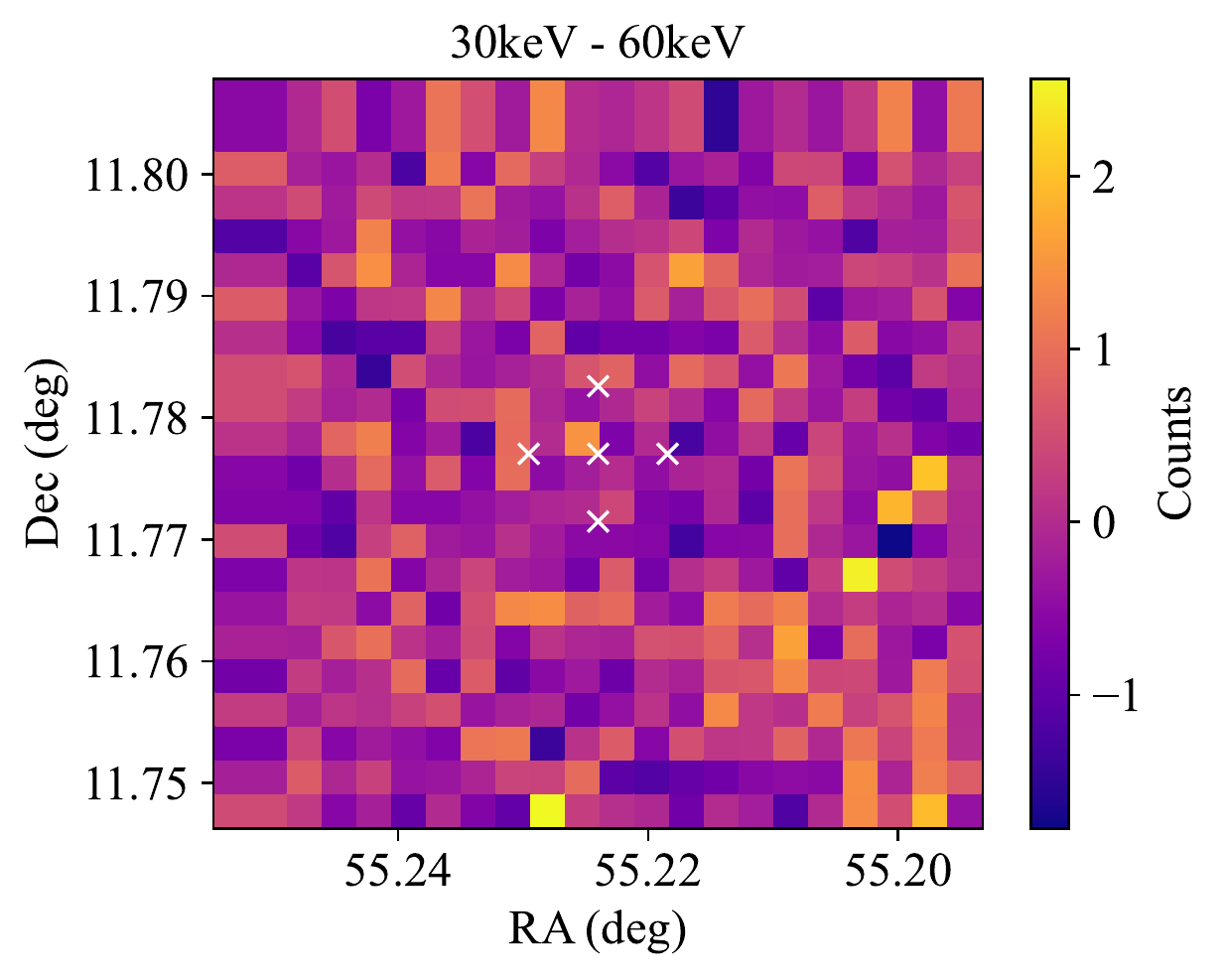}
    \caption{Stacked, background-subtracted photometry of \sn{} in four energy ranges obtained using a grid of $10''\times10''$ squared apertures. Each image has sides of  $200''$ and is oriented with North to the top and East to the left. White crosses guide the eye by indicating the expected location of \sn{} and $20''$ separations in both directions.}
    \label{fig:sn_stamps}
\end{figure}

We performed photometry using a circular aperture with radius $r=40''$, which is large enough to account for small ($<10''$) offsets possibly present in the astrometric calibration. The background region was chosen in the same detector where \sn\ was expected to be found and had a radius of $r = 100''$. 
Our \nustar{} observations did not reveal any source at the location of \sn{} (Fig.\,\ref{fig:sn_stamps}). 

We focused our analysis in the 10--30\,keV range because the SN is expected to be optically thick at energies below $\sim 10$\,keV (see \S\ref{sec:discussion}) and too faint to be detectable at energies above $\sim 30$\,keV because of the lower \nustar{} effective area. We obtained an upper limit of $2.4\times 10^{-4}$\,counts\,s$^{-1}$.
This was calculated as $3\times \sqrt{B_{\rm tot}}/t_{\rm{exp}}$, where $B_{\rm tot}$ is the total background from both epochs and both instruments normalized to the source aperture and $t_{\rm{exp}}$ is the total exposure time. 

\begin{figure*}
    \centering
    \includegraphics[width=0.9\textwidth]{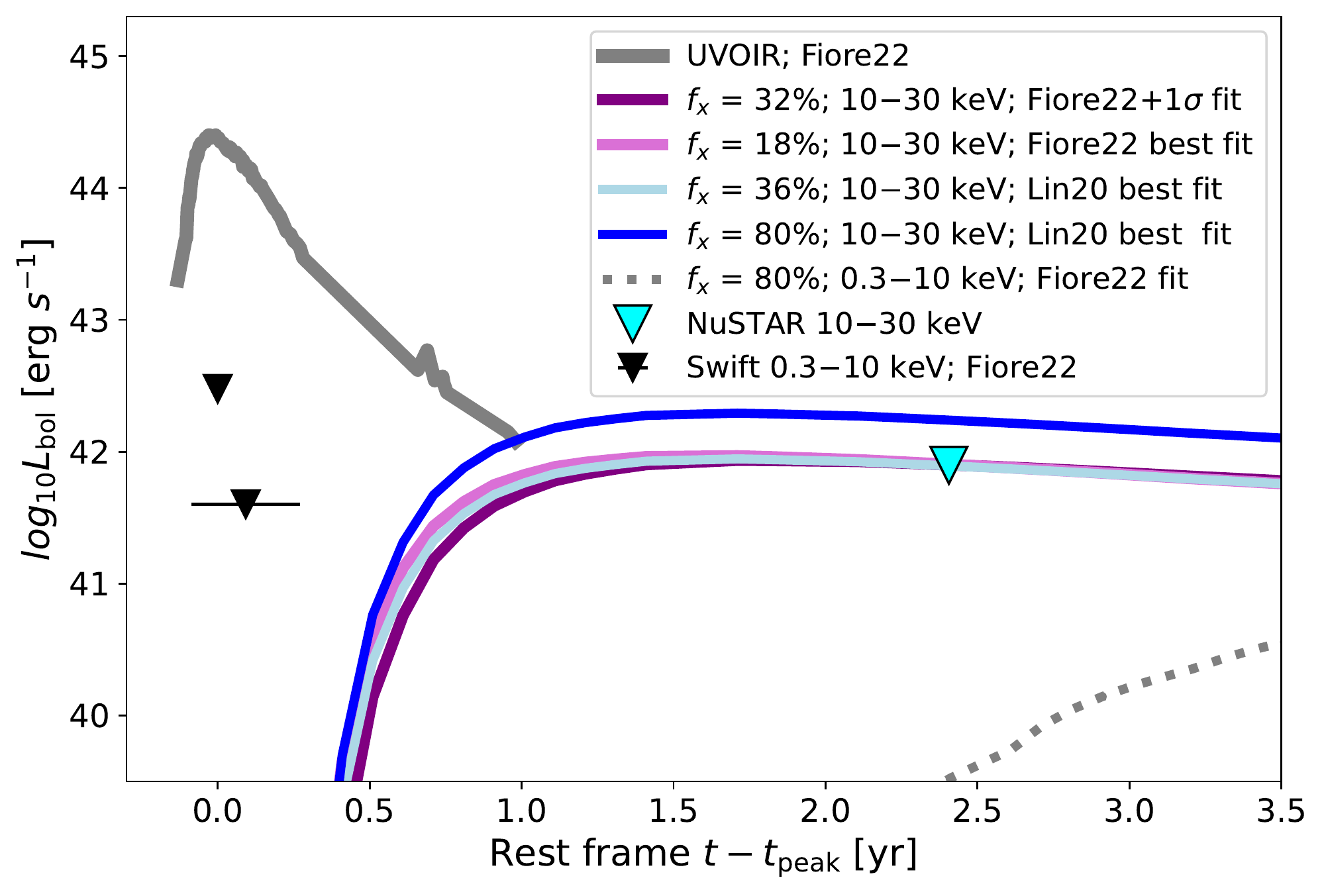}
    \caption{The \nustar{} upper limit in the 10--30\,keV range is shown (cyan triangle; $L_{\rm bol} < 7.9 \times 10^{41}$\,erg\,s$^{-1}$) along with the pseudo-bolometric light curve \citep[grey line;][]{Fiore2022} and $Swift$ upper limits \citep[black triangles;][]{Fiore2022}. The X--ray emission expected from magnetar spin-down is presented using the magnetic field and ejecta mass resulting from the fit performed by \cite{lin20_2018hti} and \cite{Fiore2022} for several values of $f_{\rm X}$. We represented the model using the best fit values for $B$ and $M_{\rm ej}$ for \cite{Fiore2022} (Fiore22 best fit) as well as the best values plus the 1$\sigma$ uncertainty in the parameter estimation (Fiore22$+1\sigma$ fit).  Models brighter than the \nustar\, upper limit are excluded by our observations (see \S\ref{sec:discussion}), for example the model that assumes the \cite{lin20_2018hti} best fit parameters and a large $f_X = 80\%$ (blue solid line). The emission expected in the 0.3--10\,keV range, in which the ejecta should still be optically thick, is represented with a dotted line for comparison.}
    \label{fig:lc}
\end{figure*}

Assuming a power-law model with photon index $\Gamma = 2$ normalized to match the expected count rate, this corresponds to an unabsorbed flux of $F_{\rm{10-30keV}} = 5.4\times 10^{-14}$\,erg\,cm$^{-2}$\,s$^{-1}$ in the 10--30\,keV range, or an upper limit of $4.8 \times 10^{41}$\,erg\,s$^{-1}$ at a luminosity distance of 271\,Mpc \citep{Fiore2022}. We defined a multiplicative absorption factor $e^{-\tau(E)}$ in \texttt{XSpec}, with $\tau(E)$ defined as in Eq.\,\ref{eq:3}. The resulting flux with absorption from the ejecta taken into account is $F_{\rm{10-30keV}} = 9.0\times 10^{-14}$\,erg\,cm$^{-2}$\,s$^{-1}$, which leads to an upper limit of $7.9 \times 10^{41}$\,erg\,s$^{-1}$ at a luminosity distance of 271\,Mpc. The result is shown in Fig.\,\ref{fig:lc}. 

We note that this luminosity upper limit is independent of the assumption on the neutral hydrogen column density $N_{\rm H}$ in the interstellar medium of the host galaxy and the Milky Way, because the 10--30\,keV flux only gets significantly attenuated by the interstellar medium for $N_{\rm H}\gtrsim 10^{24}\rm\, cm^{-2}$ \citep{wilms00_ISM_absorption}, which is unlikely given that the source is observed in the optical band.

\begin{figure}
    \centering
    \includegraphics[width=0.49\columnwidth]{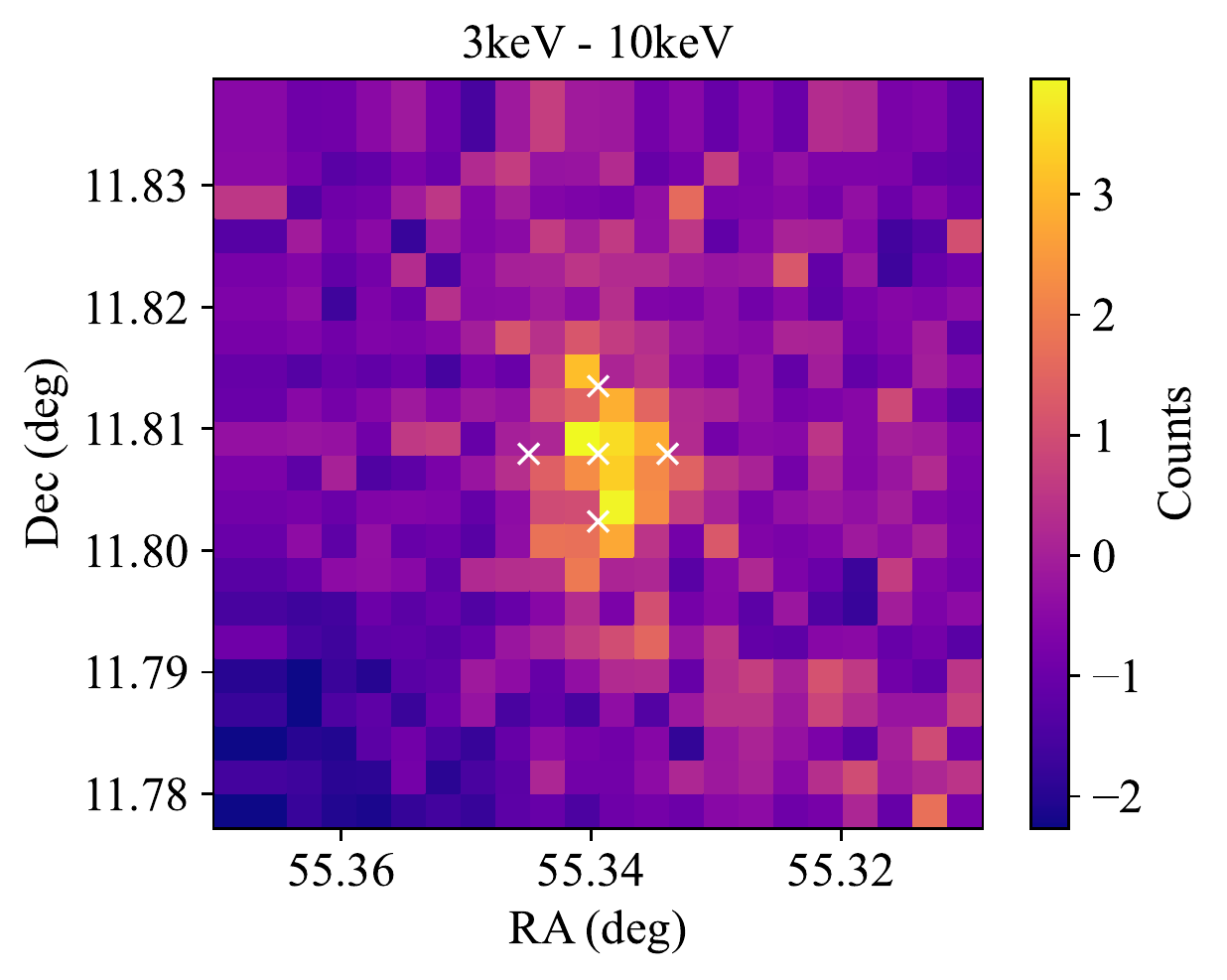}
    \includegraphics[width=0.49\columnwidth]{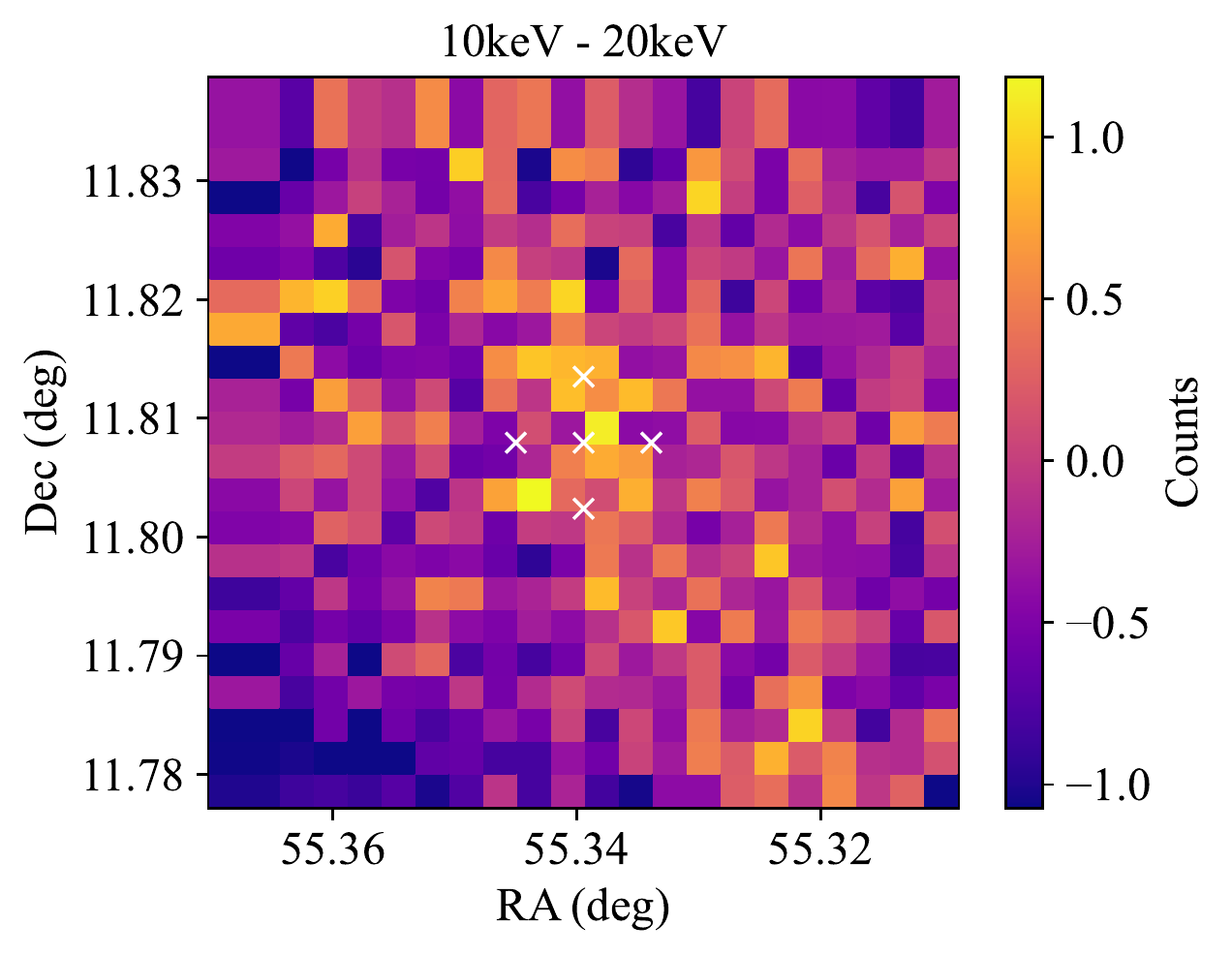}
    \includegraphics[width=0.49\columnwidth]{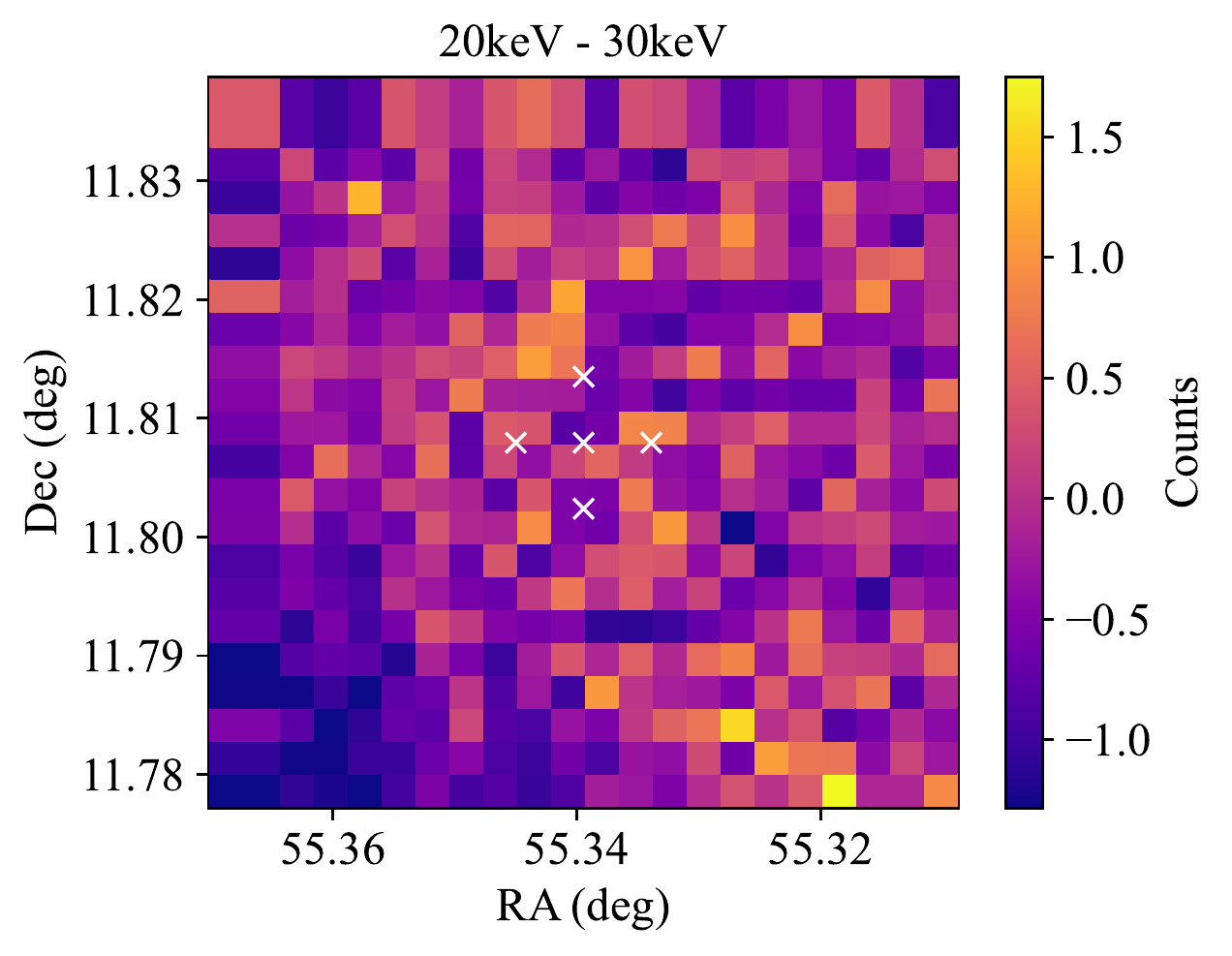}
    \includegraphics[width=0.49\columnwidth]{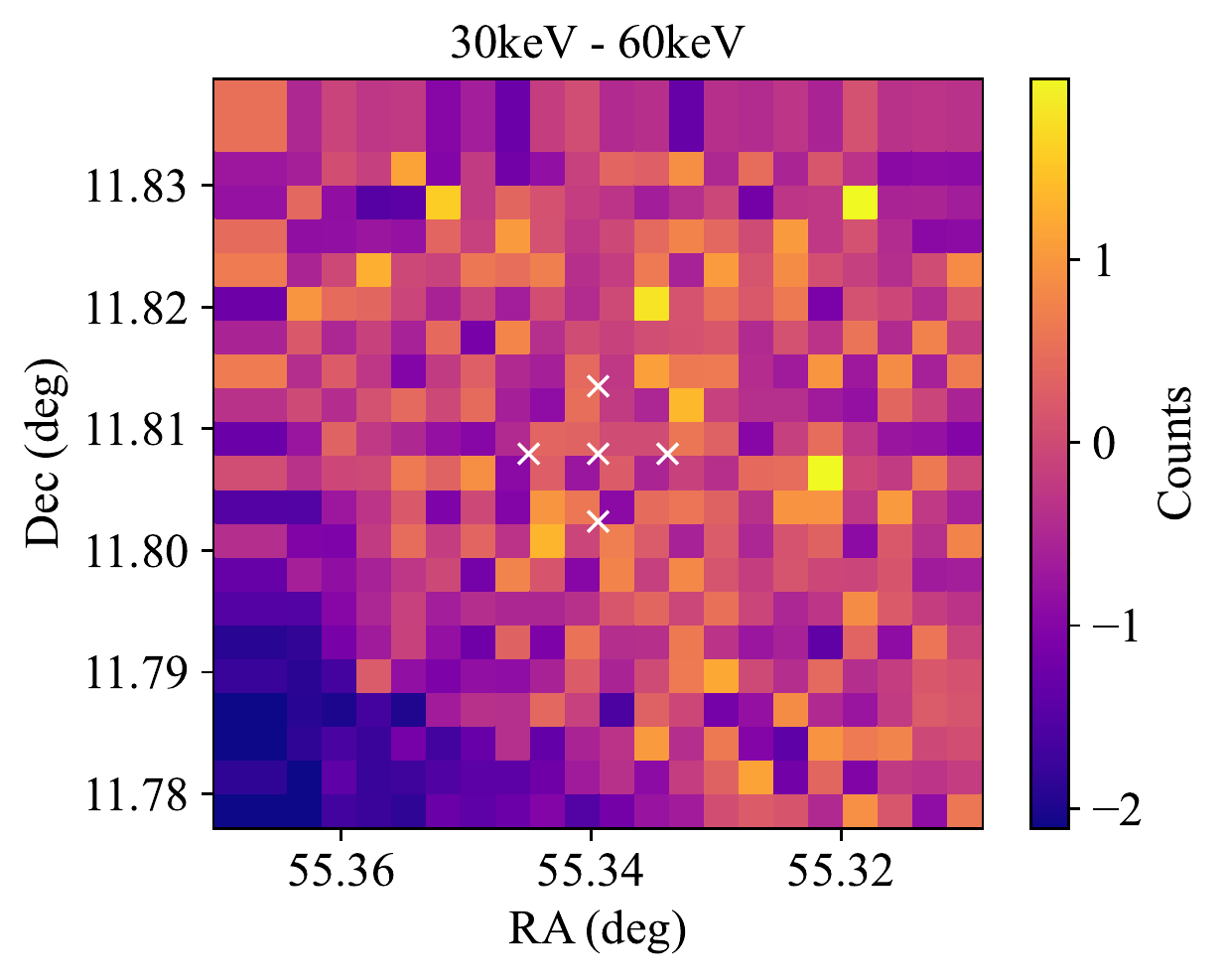}
    \caption{Images of an X--ray source serendipitously observed in the field, obtained as those in Fig.\,\ref{fig:sn_stamps}.}
    \label{fig:field_source_stamps}
\end{figure}

During the analysis, a new source was found serendipitously in the field (Fig.\,\ref{fig:field_source_stamps}).  The source is located at coordinates RA\,$=$\,03$^{\rm{h}}$41$^{\rm{m}}$21$^{\rm{s}}$; Dec\,$=$\,+11${\degree}$48$'$29$''$ ($\sim 30''$ error radius). One cataloged AGN candidate, WISEA J034122.85+114833.2, is located $27.6''$ away from the \nustar{} position, close to the edge of the error region, so an association between the two sources cannot be excluded. Follow-up observations to determine its nature are planned. Since this source was found on a different detector than \sn, its presence did not affect our SN analysis.

\begin{figure}[t]
    \centering
    \includegraphics[width=0.97\columnwidth]{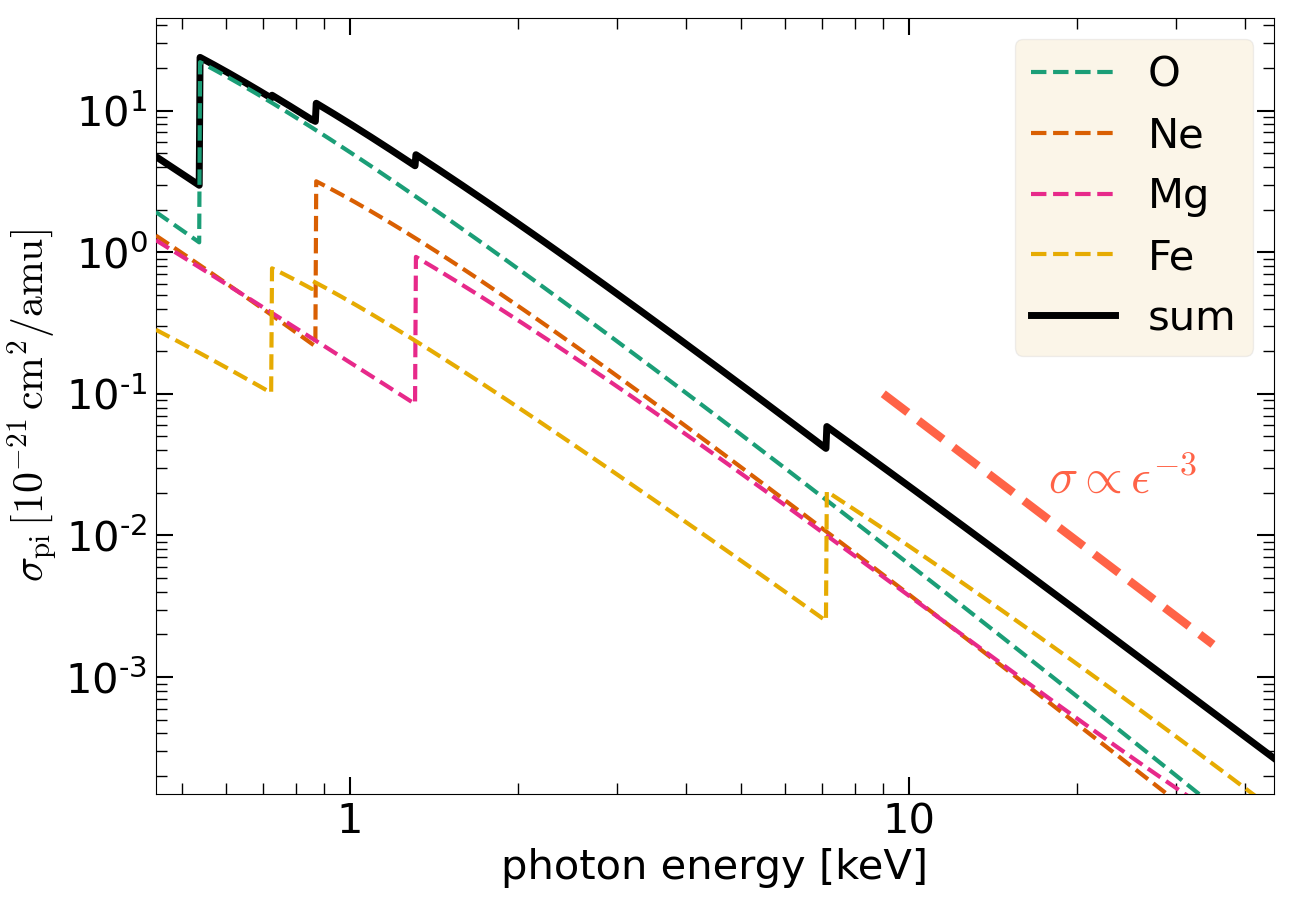}
    \caption{Photo-ionization opacity (in units of $\rm cm^2$ per atomic mass unit) for a model ejecta with abundance mass ratio of $\rm O:Ne:Mg:Fe=0.76:0.15:0.07:0.02$. The thick black line shows the total opacity and the dashed lines show contributions from different species. For this composition, the bound-free opacity is dominated by O and Fe. }
    \label{fig:opacity}
\end{figure}

\section{Discussion}
\label{sec:discussion}

To obtain a good fit to the late-time ($>100$ days) optical light curve, magnetar models have been developed by \cite{chen15_PTF12dam_leakage, wang15_energy_leakage} to allow the magnetar spin-down luminosity to leak out of the ejecta. Such a model has been used to fit the multi-color light curves of a large number of SLSNe-I, and the authors in \cite{nicholl17_fitting_results, lin20_2018hti} provided the Bayesian posteriors of the magnetic field strength $B$ and ejecta mass $M_{\rm ej}$.
The characteristic ejecta velocity $v_{\rm ej}$ is approximated by the photospheric expansion velocity inferred from the absorption line widths \citep{liu17_ejecta_speed}, and typical values are $v_{\rm ej}\simeq 1\times10^4\mr{\, km\, s^{-1}}$. 
Using the same framework as in \cite{nicholl17_fitting_results}, we expect the magnetar heating luminosity at late time ($t\gtrsim 1\rm\, yr$) to be
\begin{equation}
  \label{eq:1}
  L_{\rm mag} \simeq
  3.4\times10^{42}\mr{\,erg\,s^{-1}}\, (B/10^{14}\mr{G})^{-2}
  (t/\mr{yr})^{-2}. 
\end{equation}

In the hard X-ray band, the absorption opacity of the SN ejecta is dominated by bound-free ionization of K-shell electrons. Without a generally accepted progenitor model for SLSNe-I \citep[see][for a review of the proposed models]{moriya18_SLSN_review}, their ejecta abundances are only weakly constrained by spectroscopic observations so far. Throughout the evolution, SLSN-I spectra show absorption and emission lines of C, O, Na, Mg, Si, Ca, Fe in low ionization states, and their nebular spectra resemble those of Type Ic SNe such as SN1998bw \citep{jerkstrand17_SLSN_nebular_spectra, nicholl21_SLSN_review}. Models \citep{dessart12_SLSN_spectral_fit, mazzali16_SLSN_spectral_fit, jerkstrand17_SLSN_nebular_spectra} that produce a reasonable fit to the observed spectra generally have ejecta masses dominated by O, Ne, Mg (products of C-burning). Since the magnetar models that fit the lightcurves usually do not require heating from $\rm {}^{56}Ni$ decay, the mass of the Fe-group elements may be small (although it is only weakly constrained). Thus, we expect the bound-free opacity in the hard X-ray band to be dominated O, Ne, Mg as well as Fe if the explosions produce a substantial $\rm {}^{56}Ni$ mass $M_{\rm Ni}\gtrsim 0.1M_\odot$. To estimate the ejecta opacity, we take a fiducial abundance profile based on C-burning ashes of a $25M_\odot$ massive star from \citet{jerkstrand17_SLSN_nebular_spectra, woosley07_massive_star_evolution} and then added 2\% of iron (corresponding to $M_{\rm Ni}$ of the order $0.1M_\odot$) to obtain the final mass ratio of $\rm O:Ne:Mg:Fe=0.76:0.15:0.07:0.02$. A higher Fe mass fraction would give larger bound-free opacity in the hard X-ray band and hence our constraint on the hard X-ray luminosity from the central engine would be weaker.

The bound-free opacity according to our fiducial abundance profile, as computed using the analytic fits for the photoionization cross-sections for neutral\footnote{The photo-ionization cross-sections for K-shell electrons at energies much above the threshold depend weakly on the ionization states of outer shell electrons.} atoms \citep{verner95_inner_shells, verner96_outer_shells}, is shown in Fig. \ref{fig:opacity} and is analytically given by (for photon energy $E\gtrsim10\rm\, keV$)
\begin{equation}
    \kappa_{\rm bf}(E) \simeq 11\, \left(E\over {\rm 10\,keV}\right)^{-3} \, \mr{cm^2\,g^{-1}}.
\end{equation}

Therefore, the absorption optical depth of the ejecta when it is at a radius $r = v_{\rm ej} t$ at time $t$ since explosion is given by
\begin{equation}
  \label{eq:3}
\begin{split}
  &\tau(E) \simeq {3\kappa_{\rm bf} M_{\rm ej}\over 4\pi (v_{\rm ej}t)^2} \\
  &\simeq 4.5\, \left(E\over 10{\rm\, keV}\right)^{-3} {M_{\rm ej}\over 5M_\odot} \left(v_{\rm
    ej}\over 10^4\,\mr{km\,s^{-1}}\right)^{-2} \left(t\over 2.4\,\mr{yr}\right)^{-2}.
\end{split}
\end{equation}

This means that the ejecta may be optically thick to soft X--rays for decades, but hard X-rays $\gtrsim 10\,$keV may escape a few years after the explosion, while the magnetar luminosity is still high (see Eq.\,\ref{eq:1}). \footnote{It should be noted that Eq.\,\ref{eq:3} minimizes the optical depth by assuming that the ejecta are uniformly distributed in a sphere. If the ejecta were in a shell, the optical depth would be larger. If the distribution was clumpy it could have holes, or even a higher line-of-sight opacity.}

To constrain the fraction of the missing energy in the hard X--rays, we assumed that a fraction $f_{\rm X}=L_{\rm {10\mbox{--}30\,keV}}/L_{\rm mag}$ of the bolometric luminosity expected from the spinning down magnetar is emitted as a power-law in the 10--30\,keV band, probed by \nustar, and that the photon index is $\Gamma = 2$. From a theoretical point of view, this fraction $f_{\rm X}$ is highly uncertain. The radiation spectrum of the magnetar wind nebula depends on how particles are accelerated near the wind termination shock (located at the inner edge of the SN ejecta), the competition between synchrotron and inverse-Compton cooling of the shock-accelerated particles, as well as photon-photon pair production \citep{metzger14_ionization_breakout, vurm21_SLSN_gamma-ray}.

No X--ray flux from the SLSN central engine was revealed by these \nustar\ observations, however we can constrain $f_X$.  Using the magnetic field and ejecta mass inferred by  \cite{lin20_2018hti}, $B=1.8\times 10^{13}$\,G and $M_{\rm{ej}} = 5.8$\,M$_{\odot}$, the \nustar\ upper limit of $7.9 \times 10^{41}$\,erg\,s$^{-1}$  leads to a constraint of $f_{\rm X} < 0.36$ (Fig.\,\ref{fig:lc}). 
\cite{Fiore2022} obtained $B=1.3^{+0.3}_{-0.2} \times 10^{13}$\,G, $M_{\rm{ej}} = 5.2^{+1.4}_{-0.9}$\,M$_{\odot}$, and a velocity $v_{\rm ej}=0.95\times 10^4\mr{\, km\, s^{-1}}$. The best fit values lead to a constraint of $f_{\rm X} < 0.18$, more stringent that using the parameters obtained by \cite{lin20_2018hti}.  When accounting for the 1$\sigma$ uncertainty in the parameter estimation \citep{Fiore2022}, upper limits of $f_{\rm X} < 0.32$ ($B=1.6 \times 10^{13}$\,G, $M_{\rm{ej}} = 6.6$\,M$_{\odot}$) and $f_{\rm X} < 0.11$ ($B=1.1 \times 10^{13}$\,G, $M_{\rm{ej}} = 4.3$\,M$_{\odot}$) can be inferred. In conclusion, an upper limit of $f_{\rm X} < 0.36$ is conservative for the parameters taken from both \cite{Fiore2022} and \cite{lin20_2018hti}. Our constraints on $f_X$ are summarized in Tab.\,\ref{tab:limits}, where upper limits obtained without the absorption factor are also reported.

It is possible that only a fraction $f_{\rm r}<1$ of the spin-down power of the magnetar is converted into radiation which then heats the supernova ejecta --- the rest of the spin-down power may go into pair creation, kinetic energy of the expansion, and escaping Poynting flux. 
If this is the case, then the heating luminosity is lower than in our Eq. (\ref{eq:1}), which is used in the fitting models under the assumption that the heat and spin-down luminosities are equal, by a factor of $f_{\rm r}^{-1}$. 

Our final constraint is based on the heating luminosity $L_{\rm mag}(t=2.4\rm\, yr)$, which is a power-law ($t^{-2}$) extrapolation from the heating luminosity required to reproduce the optical lightcurve at earlier epochs.
 
In reality, if the bolometric radiative efficiency is time dependent in the first few years $f_{\rm r}(t)$, this time dependence should be included in the extrapolation. It is beyond the scope of this work to calculate $f_{\rm r}(t)$ for the magnetar wind in the first few years, but we note that in the model presented in \citet{vurm21_SLSN_gamma-ray}, relativistic particles injected by the pulsar wind are in the fast cooling regime in the first few decades (see their Eq. 23), so the radiative efficiency $f_{\rm r}$ remains roughly constant in the first few years. It is straightforward to scale our constraint on $f_{\rm X}$ to alternative models where the heating luminosity has a different time dependence than the $t^{-2}$ power-law.

\begin{table*}[t]
\centering
    \begin{tabular}{cccccc}
    \hline \hline
       B & $M_{\rm ej}$ & Comment & Reference & Upper Limit & Upper Limit\\
       ($10^{13}$\,G) & (M$_{\odot}$) &  & & unabsorbed & absorbed\\
       \hline
       1.8 & 5.8 & best fit & \cite{lin20_2018hti} & $< 22\%$ & $<36\%$ \\
       1.3 & 5.2 & best fit & \cite{Fiore2022} & $< 11\%$ & $<18\%$ \\
       1.6 & 6.6 & best fit + 1$\sigma$ & \cite{Fiore2022} & $< 19\%$ & $<32\%$ \\
       1.1 & 4.3 & best fit - 1$\sigma$ & \cite{Fiore2022} & $< 7\%$ & $<11\%$ \\
    \hline
    \end{tabular}
    \caption{Upper limit on the fraction $f_{\rm X}$ of the bolometric luminosity expected from the spinning down magnetar, emitted in the 10--30\,keV band, assuming negligible absorption (fifth column) and absorption with optical depth $\tau(E)$ as in Eq.\,\ref{eq:3} (sixth column). The upper limits on the 10--30\,keV band that we obtained are $7.9 \times 10^{41}$\,erg\,s$^{-1}$ and $4.8 \times 10^{41}$\,erg\,s$^{-1}$ for the absorbed and unabsorbed case, respectively. The first column lists the magnetic field and the second column the ejecta mass obtained by model fitting by \cite{lin20_2018hti} or \cite{Fiore2022} (fourth column).
    }
    \label{tab:limits}
\end{table*}

\vspace{-0.4cm}

\section{Conclusion}
\label{sec:conclusion}
We conducted \nustar\ observations of hydrogen-poor \sn with the goal of measuring the fraction of luminosity emitted in the hard X--rays assuming a magnetar central engine. Bound-free processes make the ejecta optically thick to soft X--rays, but we estimated that flux should leak out at energies $\gtrsim 15$\,keV after $2.4$\,yr from the explosion time of \sn. However, \nustar\ observations resulted in an upper limit on the flux of $F_{10-30\rm{keV}} = 7.9 \times 10^{41}$\,erg\,s$^{-1}$ ($F_{10-30\rm{keV}} = 4.8 \times 10^{41}$\,erg\,s$^{-1}$ without accounting for absorption by the ejecta) at a luminosity distance of 271\,Mpc. 

These results imply that the fraction of hard X--rays (10--30\,keV range) is $f_{\rm{X}} \lesssim 36\%$ of the bolometric luminosity expected from a magnetar spin-down, considering the values obtained by the less stringent model for of magnetic field $B$ and ejecta mass $M_{\rm{ej}}$ \citep[][see \S\ref{sec:discussion}]{Fiore2022}, $f_{\rm{X}} \lesssim 11\%$ for the most optimistic model, in which the SLSN is expected to be brighter. Our constraints on the high-energy spectrum of SLSNe provide important guidance in future modeling of magnetar wind nebulae, provided that they are indeed the central engine of these events.

\begin{acknowledgments}
We thank the anonymous referee for the constructive review of the manuscript. We thank Cole Miller and Simone Dichiara for the useful discussions.
This project was funded by NASA Grant Number 80NSSC22K0063. WL was supported by the Lyman Spitzer, Jr. Fellowship at Princeton University. JH acknowledges support from an appointment to the NASA Postdoctoral Program at the Goddard Space Flight Center, administered by the ORAU through a contract with NASA.
\end{acknowledgments}

\vspace{5mm}
\facilities{\nustar}

\software{astropy \citep{2013A&A...558A..33A,2018AJ....156..123A},  
          HEAsoft \citep{2014ascl.soft08004N}, XSpec \citep{Arnaud1996xspec}
          }

\bibliographystyle{aasjournal}
\bibliography{references_slsn}



\end{document}